\documentclass[
    ,final            
  ]
  {aipproc}

\layoutstyle{6x9}

\def\ergps      {$\rm ergs \ {\rm s}^{-1}$ }

\def\green{f_{_{\rm G}}}
\def\columntotal{\Phi_\epsilon}

\def\colrad{r_0}

\def\tauperp{\tau_\perp}

\def\sigpar{\sigma_{_{||}}}
\def\sigperp{\sigma_\perp}
\def\sigbar{\overline\sigma}

\newcommand{\gapprox}   {\lower.4ex\hbox{$\;\buildrel >\over{\scriptstyle\sim}\;$}}
\newcommand{\lapprox}   {\lower.4ex\hbox{$\;\buildrel <\over{\scriptstyle\sim}\;$}}
\newcommand{\begeq}     {\begin{equation}}
\newcommand{\fineq}     {\end{equation}}

\newcommand{\Tmound}    {T_{\rm th}}

\newcommand{\sig}       {\sigma_{_{\rm T}}}

\newcommand{\Msun}      {\mbox{$\,M_{\mathord\odot}$}}

\newcommand{\herxone}{Her~X-1}

\newcommand{\xper}{\mbox{X\,\,Per}}

\begin{document}

\title{Comptonization and the Spectra of Accretion-Powered X-Ray Pulsars}

\classification{95.30.Jx, 97.60.Gb, 97.80.Jp, 98.35.Mp}

\keywords{Neutron stars, Accretion, X-Rays, Comptonization, 
Radiation-dominated shocks}

\author{Michael T. Wolff}{
  address={Space Science Division, 
Naval Research Laboratory, Washington, DC 20375}}

\author{Peter A. Becker}{
  address={Center for Earth Observing and Space Research,
George Mason University, Fairfax, VA 22030-4444}}

\author{Kenneth D. Wolfram}{
  address={Center for Earth Observing and Space Research,
George Mason University, Fairfax, VA 22030-4444}
  ,altaddress={Space Science Division,
Naval Research Laboratory, Washington, DC 20375} }

\begin{abstract}
Accretion-powered X-ray pulsars are among the most luminous X-ray
sources in the Galaxy. However, despite decades of theoretical and
observational work since their discovery, no satisfactory model for the
formation of the observed X-ray spectra has emerged. In this paper, we
report on a self-consistent calculation of the spectrum emerging from a
pulsar accretion column that includes an explicit treatment of the bulk
and thermal Comptonization occurring in the radiation-dominated shocks
that form in the accretion flows. Using a rigorous eigenfunction
expansion method, we obtain a closed-form expression for the Green's
function describing the upscattering of monochromatic radiation injected
into the column. The Green's function is convolved with bremsstrahlung,
cyclotron, and blackbody source terms to calculate the emergent photon
spectrum. We show that energization of photons in the shock naturally
produces an X-ray spectrum with a relatively flat continuum and a
high-energy exponential cutoff. Finally, we demonstrate that our model
yields good agreement with the spectra of the bright pulsar \herxone\
and the low luminosity \xper.
\end{abstract}

\maketitle

\section{Introduction}

Accreting X-ray pulsars were discovered in the late 1960s by rocket
flights from White Sands in New Mexico \citep{cmr+67,ggk+71}. Since that
time the number of known X-ray pulsars of all types has grown to over 100. 
The prevailing model for accreting X-ray pulsars consists of a neutron star 
orbiting a normal stellar companion that is losing matter via either 
Roche lobe overflow or a stellar wind. Accreting X-ray pulsar spectra are
characterized by a power-law dependence on energy in the range above
$\sim 3$ keV with a quasi-exponential cutoff at higher energies,
typically near 20-40~keV. For example, see the X-ray spectrum
of 4U11626-67 reported by \citet{orla05}, or the multiple source spectra
reported by \citet{chr+02}. Cyclotron features have also been observed
in many sources \citep{chr+02}.

Comptonization is known to play an important role in the X-ray pulsar
spectral formation process. This is a general result, based on the
overall spectral shape (a power-law) and the fact that Comptonization
tends to result in a power-law dependence with energy in many accreting
compact object sources. The usual method of characterizing accreting
X-ray pulsar spectra is to fit the broad-band X-ray spectra using a
number of ad-hoc functions such as power-laws, Gaussian emission line
features, Gaussian or Lorentzian absorption features, and various types
of quasi-exponential high-energy cutoffs. However, most of these
functional forms have little physical motivation beyond the fact that
they ``look like the spectra'' and they are easy to incorporate into
spectral fitting programs (e.g., XSPEC). The real physical parameters of
the source (e.g., accretion rate, accretion region size, shock height,
plasma temperature, etc.) are not directly connected with the ad hoc
parameters and are in fact much harder to determine.

Three principal models have been put forward to explain the energy
spectra of accreting X-ray pulsars: the gas-mediated collisionless
shock model \citep{lr82}; the Coulomb collisional stopping model
\citep{mhkg83,msw87,mws89}; and the radiation-dominated flow model
\citep{davi73,akl87}. The collisionless shock model and the Coulomb
collisional model can only be applied in the case of low-luminosity
accretion onto neutron stars. In such models, the effects of radiation
pressure are assumed to be small, and the flow impinges directly onto
the neutron star surface (in the absence of other effects; see below).
These models tend to produce pencil-beamed emission patterns because the
radiation escapes primarily through the top of the neutron star
atmosphere. In high-luminosity pulsars, the dynamical structure is
expected to be dominated by the effects of radiation pressure, which
decelerates the gas to rest at the stellar surface. The
radiation-dominated inflow models of \citet{davi73} and \citet{akl87}
attempt to describe the flow dynamics across a broad range of X-ray
pulsar luminosities, all the way up to the Eddington limit.

Neither the collisionless shock model nor the Coulomb collisional model
have demonstrated good agreement with actual X-ray pulsar spectra.
\citet{mn85} compared the spectrum of \herxone\ with results obtained
using the Coulomb collisional model. The observed \herxone\ spectrum 
was not well fit by the calculated spectra. Furthermore, the applicability 
of the Coulomb collisional model to \herxone\ is questionable because the 
source luminosity is believed to be close to the Eddington limit 
($L_{\rm x} \sim 2.2 \times 10^{37}$ \ergps\ at 5 kpc), implying that 
radiation pressure is important \citep{wsh83,focs98}. For further discussion 
of Coulomb collisional stopping models see \citet{hl06} and references therein.

Insight into the photon transport in the accretion column can be
gained by estimating the optical depth to electron scattering in the
magnetic field direction, $\tau_{\parallel}$, and the scattering
optical depth across the accretion column, $\tau_{\perp}$. Assuming
Thomson scattering and a free-fall velocity profile, we can express
these quantities in cgs units using \citep[see][]{id83}
\begin{equation}
\tau_{\parallel} \sim 20 \left(L_{\rm x} \over 10^{37} \right)
\left( R_{\rm ns} \over 10^6 \right)^{5/2}
\left( M_{\rm ns} \over 1.4 \Msun \right)^{-3/2}
\left( r_0 \over 10^5 \right)^{-2} \ ,
\label{eq:tauparallel}
\end{equation}
and
\begin{equation}
\tau_{\perp} \sim 3 \left( L_{\rm x} \over 10^{37} \right)
\left( R_{\rm ns} \over 10^6 \right)^{3/2}
\left( M_{\rm ns} \over 1.4 \Msun \right)^{-3/2}
\left( r_0 \over 10^5 \right)^{-1}  \ ,
\label{eq:tauperpendicular}
\end{equation}
where $L_{\rm x}$ is the accretion luminosity, $M_{\rm ns}$ and $R_{\rm
ns}$ are the neutron star mass and radius, respectively, and $r_0$ is
the radius of the accretion region on the neutron star surface. Based on
the modeling described below, we find that $\tau_{\parallel} \sim 6.5
\times 10^{-2}$ and $\tau_{\perp} \sim 4.2 \times 10^{-3}$ for \xper.
Conversely, in the \herxone\ case we obtain $\tau_{\parallel} \sim 2.1
\times 10^4$ and $\tau_{\perp} \sim 1.4 \times 10^2$. Hence photons will
scatter many more times while escaping from the accretion flow in
\herxone\ than in \xper. This is a general result that applies to all
high-luminosity X-ray pulsars such as \herxone\ because of the larger
accretion rates compared with the low-luminosity, steep spectrum sources
such as \xper.

\section{A Radiation-Dominated Shock Model for X-Ray Pulsar Spectra}

Adopting a cylindrical, plane-parallel geometry for the accretion column
with the magnetic field in the z-direction, the Green's function
$\green(z_0,z,\epsilon_0,\epsilon)$ satisfies the modified
\citep{komp57} steady-state transport equation 
\citep[see, e.g.,][]{bw06,bb86a}
\begin{eqnarray}
v \, {\partial \green \over \partial z}
&=& {dv \over d z}\,{\epsilon\over 3} \,
{\partial \green\over\partial\epsilon}
+ {\partial\over\partial z}
\left({c\over 3 n_e \sigpar}\,{\partial \green\over\partial z}\right)
- {\green \over t_{\rm esc}}
+ {n_e \sigbar \, c \over m_e c^2} {1 \over\epsilon^2}
{\partial\over\partial\epsilon}\left[\epsilon^4\left(\green
+ k T_e \, {\partial \green\over\partial\epsilon}\right)\right]
\nonumber
\\
&+& {\dot N_0 \, \delta(\epsilon-\epsilon_0) \, \delta(z-z_0)
\over \pi r_0^2 \epsilon_0^2}
\ ,
\label{eq:radiationtransfer}
\end{eqnarray}
where $z$ is the altitude above the stellar surface, $v < 0$ is the
inflow velocity, $\dot N_0$ is the rate of injection of seed photons
with energy $\epsilon_0$ at location $z_0$, $t_{\rm esc}$ represents the
mean time photons spend in the plasma before diffusing through the walls
of the column, $\sigpar$ is the electron scattering cross section for
photons propagating parallel to the magnetic field, $\sigbar$ is the
angle-averaged cross section, and $T_e$, $n_e$, and $m_e$ denote the
electron temperature, number density, and mass, respectively. The mean
escape time is computed using $t_{\rm esc} = r_0 /w_\perp$, where
$w_\perp = c / \tauperp$ is the diffusion velocity perpendicular to the
$z$-axis, $\tauperp = n_e \, \sigperp \, \colrad$ is the electron
scattering optical thickness across the column, and $\sigperp$ denotes
the electron scattering cross section for photons propagating
perpendicular to the magnetic field. The solution for the Green's
function $\green(z_0,z,\epsilon_0, \epsilon)$ is obtained by deriving
eigenvalues and associated eigenfunctions based on the set of spatial
and energetic boundary conditions for the problem \citep[see][]{bw06}.
We define a source function $Q$ such that $\epsilon^2 \, Q(z,\epsilon)
\, d\epsilon \, dz$ gives the number of seed photons injected per unit
time in the altitude range $z$ to $z+dz$ and energy range $\epsilon$ to
$\epsilon+d\epsilon$. Once we have the analytical solution for the
Green's function, the particular solution corresponding to
bremsstrahlung, cyclotron, or blackbody source distributions can be
obtained via the integral convolution
\begin{equation}
f(z,\epsilon) = \int_0^\infty\int_0^\infty
{\green(z_0,z,\epsilon_0,\epsilon) \over \dot N_0} \ \epsilon_0^2
\, Q(z_0,\epsilon_0) \, d\epsilon_0 \, dz_0
\ .
\label{eq:convolution}
\end{equation}
In the models described here, the effects of bulk and thermal
Comptonization are treated explicitly using the transport terms in
equation~(\ref{eq:radiationtransfer}). The importance of dynamical
(bulk) versus thermal Comptonization depends on the parameter $\delta
\equiv (\alpha\sigpar / 3\sigbar)(m_e c^2 / kT_e)$, where $\alpha \sim
0.3-0.5$ describes the velocity variation as a function of the optical
depth above the stellar surface. It can be shown that $\delta$ is
essentially the ratio of the ``$y$-parameters'' for bulk and thermal
Comptonization. When $\delta$ is of order unity, the two processes are
comparable, and when $\delta \gg 1$, the bulk process dominates. Another
important spectral formation parameter is $\xi \equiv (\pi r_0 m_p
c)/(\dot M \sqrt{\sigpar \sigperp})$, where $\dot M$ denotes the
accretion rate and $m_p$ is the proton mass. We find that $\xi$ is
roughly equal to the ratio of the dynamical (accretion) timescale
divided by the timescale for the photons to diffuse through the column
walls. The condition $\xi \sim 1$ must be satisfied in order to ensure
that radiation pressure decelerates the gas to rest at the stellar
surface.

\section{Results for X-Ray Pulsar Spectra}

Using our model we can compute the theoretical spectrum emitted from an
X-ray pulsar accretion column due to Comptonized bremsstrahlung, cyclotron,
and blackbody seed photons. The theoretical phase-averaged photon count
rate spectrum, $F_\epsilon (\epsilon)$, is given by
\begin{equation}
F_\epsilon(\epsilon) \equiv {\columntotal(\epsilon) \over 4 \pi D^2}
\ , \ \ \ \ \ 
\columntotal(\epsilon) \equiv \int_0^\infty {\pi \, r_0^2 \,
\epsilon^2 \over t_{\rm esc}(z)} \, f(z,\epsilon) \, dz \ ,
\label{eq:totalspectrum}
\end{equation}
where $\columntotal(\epsilon)$ represents the vertically-integrated
escaping photon number spectrum, $f(z,\epsilon)$ is computed using
equation~(\ref{eq:convolution}), and $D$ is the distance to the source.
As a check on our results for the spectra, we confirm that the number of
photons escaping from the column per unit time is exactly equal to the
number injected, as required by our steady-state scenario.

\begin{figure}
  \includegraphics[height=.3\textheight]{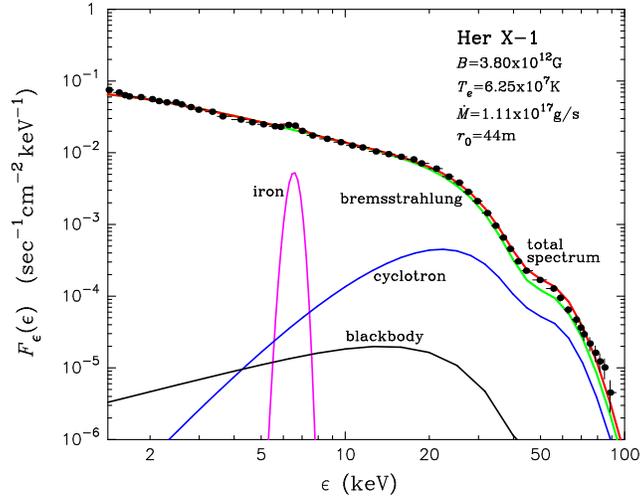}
\caption{Theoretical spectrum of \herxone\ based on our radiation-dominated,
radiative shock model (eq.~[\ref{eq:totalspectrum}]), compared with the data
reported by \citet{foc+98}.
\label{fig:herx1}}
\end{figure}

In Figure~\ref{fig:herx1} we plot the theoretical count-rate spectrum
$F_\epsilon(\epsilon)$ evaluated using equation~(\ref{eq:totalspectrum})
along with the deconvolved, phase-averaged {\it BeppoSAX} spectrum of
\herxone\ reported by \citet{foc+98}. The theoretical parameters in this
case are $\alpha=0.40$, $\xi=1.45$, $\sigperp=\sig$, $\delta=1.8$,
$B=3.80 \times 10^{12}\,$G, $\dot M=1.11 \times 10^{17}\,\rm g\,s^{-1}$,
$r_0=44\,$m, $T_e=6.25 \times 10^7\,$K, and $\Tmound=5.68 \times
10^7\,$K, where $\sig$ denotes the Thomson cross section and $\Tmound$
is the thermal mound temperature. We assume a source distance of
$D=5\,$kpc. In Figure~\ref{fig:herx1} results are plotted for the total
spectrum, as well as for the individual contributions to the observed
flux due to the Comptonization of cyclotron, blackbody, and
bremsstrahlung seed photons. The theoretical spectrum in
Figure~\ref{fig:herx1} also includes an iron emission line.

The general shape of the theoretical spectrum agrees with the
observations for \herxone\ quite well, including both the
quasi-exponential cutoff energy and the power-law slope. In the case of
\herxone, reprocessed (Comptonized) blackbody emission from the thermal
mound makes a negligible contribution to the spectrum because the radius
of the accretion column is relatively small. Due to the high temperature
of the post-shock plasma, the reprocessed cyclotron emission is
overwhelmed by reprocessed bremsstrahlung emission, which dominates the
observed spectrum.

\begin{figure}
  \includegraphics[height=.3\textheight]{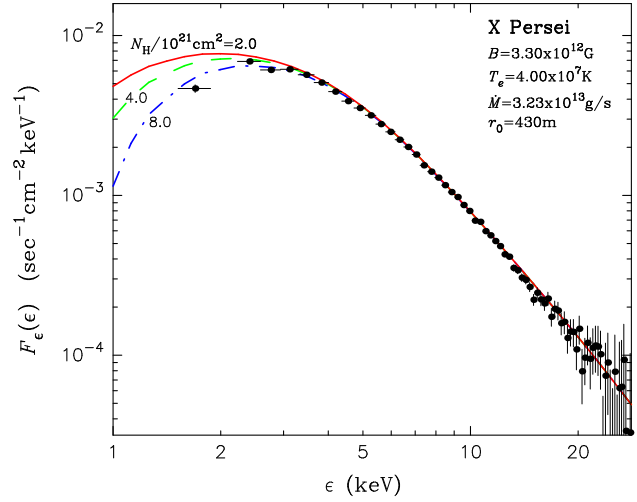}
\caption{Theoretical spectrum of \xper\ based on our radiation-dominated,
radiative shock model (eq.~[\ref{eq:totalspectrum}]), compared with the data
reported by \citet{mlpr01}. Various amounts of interstellar absorption have
been included as indicated.
\label{fig:xper}}
\end{figure}

The second source, 4U~0352+30 (\xper), is a low-luminosity ($L_{\rm x}
\sim 10^{34}{\rm \ ergs \ s}^{-1}$) X-ray pulsar with a relatively steep
spectrum. This pulsar was discussed by \citet{bw05a,bw05b} using a
pure-bulk Comptonization model, and it therefore represents an
interesting test for the thermal+bulk model developed by \citet{bw06}
and described in this paper. In Figure~\ref{fig:xper} we compare the
theoretical spectrum with archival {\it RXTE} data taken in July 1998
and reported by \citet{mlpr01}. The theory parameters in this case are
$\alpha=0.51$, $\xi=1.85$, $\sigperp=\sig$, $\delta=10.9$, $B=3.30
\times 10^{12}\,$G, $\dot M=3.23 \times 10^{13}\,{\rm g \, s}^{-1}$,
$r_0=430\,$m, $T_e=4.00 \times 10^7\,$K, $\Tmound=9.00 \times 10^6\,$K,
and $D = 0.35\,$kpc. The distance and magnetic field values are from
\citet{negu98} and \citet{chg+01}, respectively. In contrast to the case
of \herxone, our model shows that the spectrum of \xper\ is dominated by
Comptonized {\it blackbody} emission, which is due to the order of
magnitude increase in the radius of the accretion column. Reprocessed
cyclotron and bremsstrahlung radiation make a negligible contribution
to the observed spectrum for this source. The spectral results obtained
here using the thermal+bulk Comptonization model are nearly identical to
those obtained by \citet{bw05b,bw05a}, which is consistent with the fact
that the spectral formation in this source is dominated by bulk
Comptonization, as indicated by the large value of $\delta$.

\section{Conclusions}

We have developed a new analytical model describing the spectral
formation process in accretion-powered X-ray pulsars. The model includes
a rigorous treatment of both the bulk and thermal Comptonization
occurring in the radiation-dominated, radiative shock. These two types
of Comptonization influence different regions of the radiation
distribution and can explain a wide range of accretion-powered X-ray
pulsar spectra. We have shown that the theoretical spectra produced by
our model in luminous sources such as \herxone\ are dominated by
Comptonized bremsstrahlung emission, not Comptonized cyclotron emission
as was previously conjectured. On the other hand, we find that the
spectra of low-luminosity sources such as \xper\ are dominated by
Comptonized blackbody radiation. Our new Comptonization model provides
greatly improved fits to the observed spectral data when compared with
the gas-mediated collisionless shock or Coulomb collisional models.
Furthermore, the X-ray spectra produced by our model can be evaluated
using a single-pass, closed-form algorithm that does not require
numerical iteration. Our model should therefore be suitable for
incorporation into the XSPEC spectral modeling environment and we expect
to complete that work in the near future.

\begin{theacknowledgments}
The authors wish to thank Drs. Lev Titarchuk, Kent Wood, and Jean Swank
for useful discussions. This research was funded by NASA and the Office
of Naval Research.
\end{theacknowledgments}



\bibliographystyle{aipproc}   




\end{document}